\documentclass[reprint,aps,twocolumn,showpacs,superscriptaddress,groupedaddress,nofootinbib]{revtex4-2}
\usepackage{graphicx}  
\usepackage{tabularx}
\usepackage{booktabs}
\usepackage{dcolumn}   
\usepackage{bm}        
\usepackage{amssymb}   
\usepackage{amsmath}   
\usepackage{multirow}
\usepackage{hyperref}
\usepackage{braket}
\usepackage{subfigure}
\usepackage{array}
\usepackage[math]{cellspace}
\usepackage[utf8]{inputenc}
\usepackage{cleveref}
\usepackage{algorithm}
\usepackage{algpseudocode}
\usepackage{physics}
\algrenewcommand\algorithmicrequire{\textbf{Input:}}
\algrenewcommand\algorithmicensure{\textbf{Output:}}

\usepackage{color}
\usepackage[normalem]{ulem}

\usepackage{url}
\hypersetup{
bookmarksnumbered,
colorlinks=true,
linkcolor=magenta,
urlcolor=magenta,
citecolor=blue
} 
\begin{document}
\hspace{5.2in} 

\title{Stochastic automatic differentiation and the Signal to Noise problem}

\date{\today}

\author{G. Catumba}
\email{guilherme.catumba@mib.infn.it}
\affiliation{Department of Physics, University of Milano-Bicocca, Piazza della Scienza 3, 20126, Milano, Italy}
\affiliation{INFN Milano–Bicocca, Piazza della Scienza 3, I-20126 Milano, Italy}

\author{A. Ramos}
\email{alberto.ramos@ific.uv.es}
\affiliation{Instituto de Física Corpuscular (CSIC-UVEG), Edificio Institutos Investigaci\'on,
Apt.\ 22085, E-46071 Valencia, Spain}

\begin{abstract}
Lattice Field theory allows to extract properties of particles in
strongly coupled quantum field theories by studying  Euclidean vacuum
expectation values.  When estimated from numerical Monte Carlo
simulations these are typically affected by the so called Signal to
Noise  problem: both the signal and the variance decay exponentially
with the Euclidean time, but the variance decays slower, making the
signal to noise ratio to degrade exponentially fast.  In this work we
show that writing correlators as derivatives with respect to sources
and evaluating these derivatives using techniques of stochastic
automatic differentiation can eliminate completely the signal to noise
problem.  We show some results in scalar field theories, and comment
on the prospects for applicability in Gauge theories and QCD.
\end{abstract}
\maketitle

\section{Introduction}

Lattice field theory offers a first principle approach for
calculating properties of strongly coupled quantum field theories.  As
of today it offers e.g.  invaluable input to flavor
physics~\cite{FlavourLatticeAveragingGroupFLAG:2024oxs} and the
determination of the fundamental parameters of the Standard Model.

Many properties are extracted from the Euclidean time behavior of
connected two-point vacuum expectation values of some suitable local
interpolating operator  $O(x)$
\begin{equation}
  C(x_0) = \sum_{\vec x}\langle O^\dagger(x)O(0) \rangle_{\rm c}\,,
\end{equation}
taken with respect to the Euclidean path integral
\begin{equation}
  \label{eq:pathintegral}
  \mathcal Z = \int {\rm d} \phi\, e^{-S[\phi]}\,.
\end{equation}
The spectral decomposition
\begin{equation}
  \label{eq:correlator}
  C(x_0) = \sum_n |\langle 0 | O | n \rangle|^2 e^{-E_n x_0}
\end{equation}
shows how to access properties of the
Hamiltonian (i.e. the energy levels $E_n$), from the study of these
correlators. The typical example is the extraction of the ground state
energy $E_0$ from the so called `effective mass'
\begin{align}
  &m_{\rm eff}(x_0) = -\partial_{x_0}\log C(x_0)\,,  \\
  &E_0 = \lim_{x_0\to\infty} m_{\rm eff}(x_0)\,. \nonumber
\end{align}
A common problem is that these correlators, when estimated numerically
with Monte Carlo methods, show a degrading signal at increasing
Euclidean time.  To see this \cite{Parisi:1983ae,Lepage:1989hd}, one
can also apply the spectral decomposition to the variance, and
generically it can decay at large Euclidean times slower than the
signal itself.  This produces an exponentially large relative error in
the estimates of the effective masses.  This is referred as the
\textit{signal to noise} problem in lattice field theory.

Clearly this large noise at large Euclidean times prevents a naive
extraction of the energy levels.  Many works in the lattice literature
have attacked this problem.  Better interpolating operators allow to
obtain correlators that collapse on a single state at earlier
Euclidean times, before the signal in the correlation function is
lost. Used with a basis of several interpolating operators, we have
some of the most sucessfull approaches, like
distilation~\cite{HadronSpectrum:2009krc}, the Generalised Eigenvalue
Method (GEVP)~\cite{Luscher:1990ck,Blossier:2009kd} or the more recent
Prony method~\cite{Fischer:2020bgv} (or the equivalent Lanczos
method~\cite{Hackett:2024xnx}).

Note however that these methods do not aim at solving the signal to
noise problem.  Instead they allow to extract the spectrum from
smaller Euclidean times, where the numerical error in the correlator
is still acceptable. The reason behind these approaches is clear:  in
any Monte Carlo simulation the variance of the correlator,
Eq.~(\ref{eq:correlator}), is an observable itself, and therefore the
signal to noise problem \emph{must} be present in any Monte Carlo
simulation. The exponential deterioration of the signal seems
unavoidable.

Here we attack the problem from a different point of view. 
We will consider simulations of a different system, where sources at
fixed time slices are included in the action
\begin{equation}
  \mathcal Z_J = \int {\rm d} \phi\, e^{-S_J[\phi]}\, 
\end{equation}
with
\begin{equation}
  S_J[\phi] = S[\phi] - J\sum_{\vec x}\eval{O(x)}_{x_0=0}
\end{equation}
From this definition the correlator Eq.~(\ref{eq:correlator}) can be
determined from the expression
\begin{equation}
  \label{eq:corr_from_der}
  C(x_0) = \left.\frac{\partial }{\partial J}\right|_{J=0} \langle O^\dagger(x) \rangle_J\,,
\end{equation}
where the symbol $\langle \cdot \rangle_J$ represents expectation
values taken with respect to $\mathcal{ Z}_J$.  Contrary to the direct
evaluation of the correlator in Eq.~(\ref{eq:correlator}), the
variance in Eq.~(\ref{eq:corr_from_der}) depends crucially on how the
derivative is computed.  Following the ideas
of~\cite{Catumba:2023ulz}, the usual computation of the correlator,
Eq.~(\ref{eq:correlator}), can be understood from the computation of
the expectation values $\langle \cdot \rangle_{J}$ using
\emph{reweighting}
\begin{equation}
  \langle O^\dagger(x) \rangle_{J} = \frac{\langle e^{-(S_J[\phi]-S[\phi])}O^\dagger(x) \rangle}{\langle e^{-(S_J[\phi]-S[\phi])} \rangle}\,,
\end{equation}
i.e., the usual lattice estimate of two-point functions amounts to a
reweighting computation.  Reference~\cite{Catumba:2023ulz} has pointed
that a dramatic reduction in the variance of the numerical evaluation
of derivatives like Eq.~(\ref{eq:corr_from_der}) can be obtained by
using certain field transformations $\phi \to \varphi$.  The
transformed field $\varphi$, and in particular its statistical
distribution, carries the dependence with respect to $J$ and makes the
reweighing factors constant.  Moreover, for certain classes of quantum
field theories, these field transformations can be generated exactly.
Using a scalar field theory in 4 dimensions, we show that in these
cases, the signal to noise problem is completely eliminated:
\emph{correlators show a relative error that is independent of the
Euclidean time}.  The construction of approximate field
transformations can also produce a significant impact.

The paper is organized as follows.  Section~\ref{sec:stoch-autom-diff}
will introduce the notation and show how the proposal
of~\cite{Catumba:2023ulz} can be applied to the computation of
correlators.  Section~\ref{sec:some-results-scalar} will show our
results for $\lambda-\phi^4$ in 4d, and we will finally conclude with
some comments on the applicability for gauge theories and QCD in
section~\ref{sec:conclusions}.

\section{Stochastic automatic differentiation}
\label{sec:stoch-autom-diff}

In order to implement the numerical computation of deriatives we
consider the algebra of polynomials
\begin{equation}
  \tilde a(\epsilon) = \sum_{n=0}^N a_n\epsilon^n\,,
\end{equation}
truncated at some fixed order $N$.  The
addition/subtraction/multiplication/division, and in general the
evaluation of any function is the usual one, but with
$\mathcal O(\epsilon^{N+1})$ terms and higher neglected\footnote{As a
concrete example, for the choice $N=2$, and
$\tilde a(\epsilon) = 1+\epsilon$ and
$\tilde b(\epsilon) = 2+\epsilon^2$, we have
  \begin{displaymath}
    \tilde a(\epsilon) \times \tilde b(\epsilon) = 2 + 2\epsilon + \epsilon^2\,,
  \end{displaymath}
  and
  \begin{displaymath}
    \frac{1}{\tilde a(\epsilon)} = 1 - \epsilon + \epsilon^2º,.
  \end{displaymath}
}
(see~\cite{Catumba:2023ulz} for more details and examples).  Note that
this algebra is formally, and automatically implemented numerically as
a series of operations between the coefficients of the polynomials,
without any need to fix a value for $\epsilon$.

If one considers a function of a real variable
$f(x)$, and evaluates $f(\tilde x(\epsilon))$ using as argument the
truncated polynomial $\tilde
x(\epsilon) = x_0 + \epsilon$ and the operations of
the algebra of truncated polynomials, one obtains as result a
truncated polynomial
\begin{equation}
  \tilde f(\epsilon) = f(\tilde x(\epsilon)) = \sum_{n=0}^N f_n\epsilon^n\,.
\end{equation}
By the Taylor theorem the coefficients are proportional to the
derivatives of the function at $x_0$
\begin{equation}
  f_n = n!\, \eval{\frac{{\rm d} ^n }{{\rm d} x^n}}_{x_0} f(x)\,.
\end{equation}
This is just \emph{forward mode automatic differentiation} (see for
example~\cite{enwiki:1250217779} and
\url{https://igit.ific.uv.es/alramos/formalseries.jl} for a numerical
implementation).  

For convenience in what follows we will add a tilde over quantities
that are truncated polynomials, and will omit the argument $\epsilon$ (i.e. 
we write $\tilde a = a_0 + a_1\epsilon + \dots$).

\subsection{Automatic differentiation and correlators}

Consider an Euclidean field theory with partition function
\begin{equation}
  \mathcal Z = \int {\rm d} \phi\, e^{-S[\phi]}\, \,.
\end{equation}
For simplicity we will consider real fields and the computation of the
connected two point function of the fundamental field $\phi$ summed
over the spatial directions (projection to zero momentum)
\begin{equation}
        \label{eq:corr}
  C(y_0-x_0) = \sum_{\vec y, \vec x}\langle \phi(y)\phi(x) \rangle - \langle \phi(x) \rangle\langle \phi(y) \rangle
\end{equation}
Note that this correlator can be obtained as the derivative of a
one-point function
\begin{equation*}
  C(x_0) = \eval{\frac{\partial }{\partial J}}_{J=0} \langle \phi(x) \rangle_J\,,
\end{equation*}
after adding a source at zero euclidean time to the action
\begin{equation}
  \label{eq:action_source}
  S_J[\phi] = S[\phi] + J\sum_{\vec x}\eval{\phi(x)}_{x_0=0}\,.
\end{equation}

Automatic differentiation can be used to determine such expectation
values by introducing a fictitious source
\begin{equation}
  \tilde J = 0 + \epsilon\,.
\end{equation}
(i.e. the source is a truncated polynomial). It is straightforward to
check that if we define the action as in Eq.~(\ref{eq:action_source})
with $J$ replaced by $\tilde J$, the correlator in Eq.~(\ref{eq:corr})
can then be obtained via reweighing as the linear order in $\epsilon$
\begin{equation}
  \label{eq:rw_trunc}
  \frac{\sum_{\vec y}\langle e^{-\Delta \tilde S[\phi]} \phi(y)\rangle }{ \langle e^{-\Delta \tilde S[\phi]} \rangle } =
  \sum_{\vec y}\langle \phi(y) \rangle + C(y_0)\, \epsilon\,.
\end{equation}
Here
\begin{equation}
  \Delta \tilde S[\phi] = \tilde S_{\tilde J}[\phi] - S[\phi]\,.
\end{equation}
Equation~(\ref{eq:rw_trunc}) just shows the relation between
reweighting and the computation of derivatives w.r.t.\ a source in the
action.  Moreover, it makes clear that the usual computation of a
two-point function (i.e., the naive evalutation of $C(y_0)$) can be
seen as a reweighting procedure.

Sources have been used to avoid the computation of disconnected
diagrams in flavor singlet matrix elements~\cite{Detmold:2004kw} and
first attempted for quenched lattice QCD in~\cite{QCDSF:2012mkm}. 
In these proposals finite differences are used to estimate the numerical
derivatives. It is important to point out that using stochastic
automatic differentiation allows to obtain directly estimates of the
derivatives of interest without introducing any systematics associated
with the discretization.

\subsection{Field transformations}

If we consider a field transformation
\begin{equation}
  \label{eq:field_trans}
  \phi \to \tilde \phi = \phi + f(\phi)\, \epsilon
\end{equation}
where $f(\phi)$ is in principle an arbitrary function of the fields
such that
\begin{equation}
  \label{eq:exact}
  \tilde S_{\tilde J}[\tilde \phi] - S[\phi] - \log \left| \frac{{\rm d} \phi}{{\rm d} \tilde \phi} \right| =  \text{constant}
\end{equation}
then the change of variables transforms samples $\phi$ of $e^{-S[\phi]}$
into samples $\tilde \phi$ of $e^{-\tilde S_{\tilde J}[\tilde\phi]}$.  In
this particular case the correlator can be obtained as a
\textit{one-point function}
\begin{equation}
  \label{eq:onept}
  \sum_{\vec y}\langle \tilde \phi(y) \rangle = \langle \phi(y)\rangle + C(y_0)\, \epsilon\,.
\end{equation}
If the transformation does not exactly obey Eq.~(\ref{eq:exact}), the
correlator can still be computed, but with some modified reweighing
factors
\begin{equation}
  \label{eq:trw_factors}
  \tilde w[\tilde \phi] = \exp \left\{ -\tilde S_{\tilde J}[\tilde \phi] + \log \left| \frac{{\rm d} \tilde \phi}{{\rm d} \phi} \right| + S[\phi]  \right\}
\end{equation}
and the usual reweighing formula
\begin{equation}
  \frac{\sum_{\vec y}\langle \tilde w[\tilde \phi] \tilde \phi(y)\rangle }{ \langle \tilde w[\tilde\phi] \rangle } =
  \langle \phi(y) \rangle + C(y_0)\, \epsilon\,.
\end{equation}

In~\cite{Catumba:2023ulz} it has been shown that the variance of
derivatives such as those in Eq.~(\ref{eq:corr_from_der}) is
drastically reduced with transformations that reduce the variance of the
reweighing factors. 
In the particular case that the reweighing factors are constant, the
one point function estimate, Eq.~(\ref{eq:onept}), of the derivative
in Eq.~(\ref{eq:corr_from_der}) leads to a fairly large reduction in
the error.
It might seem that finding these transformations is difficult, but as
will see, building on the ideas of Numerical Stochastic
Perturbation Theory (NSPT)~\cite{DiRenzo:1999az} in the Hamiltonian
formulation~\cite{DallaBrida:2017tru,DallaBrida:2017pex}, one can find
these transformations for a wide class of field theories.

\subsection{Transformations and Hamiltonian flows}
\label{sec:transf-hamilt-flows}

The main algorithm to simulate Lattice gauge theories is the HMC~\cite{Duane:1987de}. 
It is based on the Hamiltonian evolution derived from
\begin{equation}
  H(\pi, \phi) = \sum_x\frac{\pi^2(x)}{2} + S[\phi]\,.
\end{equation}
where $\pi(x)$ are fictitious momenta conjugate to the field variables
$\phi(x)$. From an initial configuration $\phi$, one draws random
Gaussian momenta $\pi(x) \sim \mathcal N(0,1)$ and solves the
equations of motion (e.o.m.)
\begin{subequations}
  \label{eq:eom}
  \begin{align}
  \dot \phi(x) &= \pi(x)\,,\\
  \dot \phi(x) &= - \frac{\partial H}{\partial \phi(x)} = - \frac{\partial S[\phi]}{\partial \phi(x)} \,,
\end{align}
\end{subequations}
for a fixed (or random) time $\tau$. The field $\phi(x)$ at the end of
the trajectory (possibly corrected with a Metropolis step for the
inaccuracies caused by the 
numerical integration of the e.o.m, Eqs.~(\ref{eq:eom})) is
distributed according to the probability distribution $p[\phi]\propto e^{-S[\phi]}{\rm d} \phi$. 

As suggested in~\cite{Catumba:2023ulz}, if one promotes both the fields
and the conjugate momenta to truncated polynomials, and considers the
modified equations of motion
\begin{subequations}
  \label{eq:eompt}
  \begin{align}
  \dot {\tilde \phi}(x) &= \tilde\pi(x)\,,\\
  \dot {\tilde \phi}(x) &= - \frac{\partial \tilde S_{\tilde J}[\tilde \phi]}{\partial \tilde \phi(x)} \,,
\end{align}
\end{subequations}
the Hamiltonian flow generates fields $\tilde \phi$ whose first order
in $\epsilon$ allows to compute the linear dependence of observables
w.r.t.\ the expansion variable without any need to include reweighting
factors.  One can say that the Hamiltonian flow from
Eqs.~(\ref{eq:eompt}) exactly generates field transformations that
obey eq.~(\ref{eq:exact}).

Note that Eqs.~(\ref{eq:eompt}) can be solved numerically in a fully
explicit way (i.e.  the order $n$ at simulation time $t$ only depends
on the lower orders $0,1,\dots, n-1$ at the same simulation time $t$). 
From a computational point of view the procedure is very similar to
how NSPT is implemented.

As discussed in~\cite{Catumba:2023ulz}, to show the convergence of
the stochastic process for the non-zero order, one proceeds in a very
similar fashion as in the case of NSPT. 
It is key that the stochastic process is exploring fluctuations around
a minimum.
The matrix
\begin{equation}
  \frac{\partial^2 S}{\partial \phi(x)\partial \phi(y)} 
\end{equation}
has to be positive definite.  Many interesting theories do not have
this property (i.e.  theories with compact variables, like gauge
theories).  Also the source term $J\sum_{\vec x}\phi(x)|_{x_0=0}$ has
to be real. These restrictions imply that Hamiltonian flows, even if
they provide an exact answer to the problem of finding transformations
that obey Eq.~(\ref{eq:exact}) in some models, can not be regarded as
a generally applicable solution.  Another technical limitation is that
the accept/reject step cannot be performed to make the algorithm
exact. One has either to make sure that the results are independent of
the integration time step size $\delta t$ used to integrate the e.o.m.
Eqs.~(\ref{eq:eompt}) or perform an extrapolation $\delta t\to 0$.

\section{Some results in scalar field theories}
\label{sec:some-results-scalar}

We will use the $\lambda-\phi^4$ theory in 4 space-time dimensions as
a toy model. 
Beyond the interest in itself, this model has a very severe
signal to noise problem. At large Euclidean times, the variance of the
two-point function is constant, while the signal decreases
exponentially with the mass $m_{\rm R}$ of the particle. 
On a Monte-Carlo evaluation using $N$ samples, the error in the correlator
Eq.~(\ref{eq:corr}) grow exponentially fast with Euclidean time, while
it only decreases $\propto 1/\sqrt{N}$:
\begin{equation}
  \frac{{\rm error}[C(x_0)]}{C(x_0)} \xrightarrow[x_0\to\infty]{} \propto \frac{e^{m_{\rm R}x_0}}{\sqrt{N}} \,.
\end{equation}
On the other hand, this model shows very
little excited states contamination. 
Simple interpolators like $\phi^n(x)$ couple very strongly with
$n$-particle states, and very weakly with anything else (see for
example~\cite{Garofalo:2021ick}). 
These characteristics make the model ideal to study the signal to
noise problem (i.e. 
the variance of the correlator at large Euclidean times) separating it
from the excited states contamination. 
This is well exemplified by the free case, where there is no excited
state contamination but a severe signal to noise problem is present.

\begin{figure}
  \centering
  \includegraphics[width=0.49\textwidth]{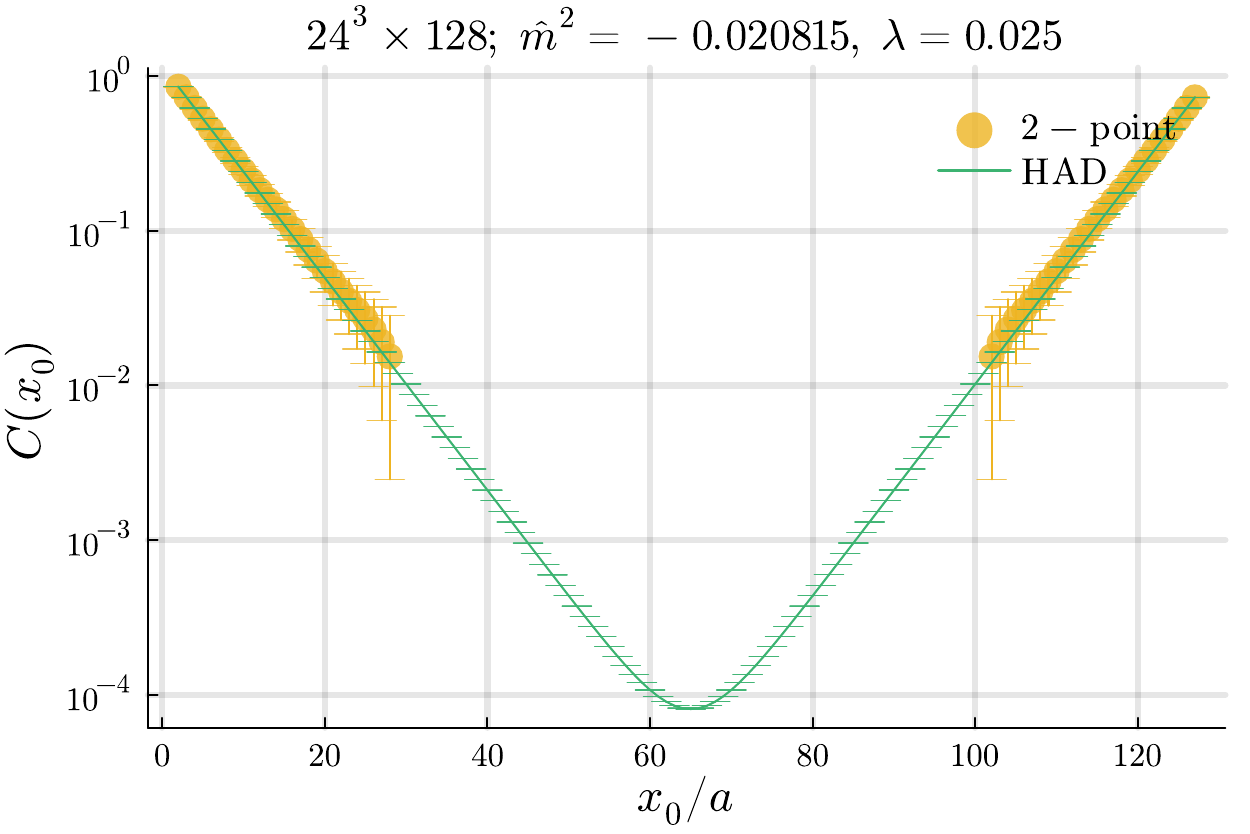}
  \caption{Correlator $C(x_0) = \sum_{\vec x}\langle \phi(x)\phi(0)
    \rangle_{\rm c}$ obtained from the usual evaluation of the two
point function Eq.~(\ref{eq:corr}) in yellow, versus the evaluation of
the one point function using a Hamiltonian approach
Eq.~(\ref{eq:onept}) an labeled HAD.  The same statistics was used for
both results, with the MC chain having around 20k HMC trajectories.
Invariance under translations was used only in the usual computation,
with an average performed over equal time separations $x_{0}-y_{0}$ to
improve the signal.}
  \label{fig:corr}
\end{figure}

The lattice action of this model is given by
\begin{align}
  \label{eq:Slatt}
  \nonumber
  S_{\rm latt}(\phi;\hat m,\lambda)=& \sum_x \left\{  \frac{1}{2}\sum_\mu[\phi(x+\hat\mu) - \phi(x)]^2\right.\\
                                             & \left. + \frac{\hat m^2}{2}\phi^2(x) + \lambda\phi^4(x) \right\}\,,
\end{align}
where the mass parameter is measured in lattice units $\hat m = am$.
We will consider simulations on a $L^3\times T$ lattice with periodic
boundary conditions in all four directions. 
This model can be simulated with the usual HMC algorithm, with the 
e.o.m. 
reading
\begin{subequations}
\begin{align}
  \dot \phi(x) &= \pi(x)\,, \\
  \dot \pi(x) &= \frac{1}{2}\sum_\mu \left[ \phi(x+\mu) + \phi(x-\mu) \right] \\
  \nonumber
               &-(4+\hat m^2)\phi(x) - 4\lambda\phi^3(x)\,.
\end{align}
\end{subequations}
After generating $N_{\rm conf}$ samples $\{\phi^{(\alpha)}
\}_{\alpha=0}^{N_{\rm conf}}$, one obtains the usual estimate of the
correlator following Eq.~(\ref{eq:corr}) from a Monte Carlo average
\begin{equation}
  C(y_0-x_0) = \frac{1}{N_{\rm conf}\times L^3}\sum_{\alpha=1}^{N_{\rm conf}}\sum_{ \vec x, \vec y} \phi^{(\alpha)}(y)\phi^{(\alpha)}(x) \,.
\end{equation}
Figure~\ref{fig:corr} shows in yellow a typical case for $C(x_{0})$
where invariance under translations was used to improve the signal.
It is apparent that the signal degrades very fast.  In fact the
variance at moderate Euclidean times becomes so large that even error
estimation is problematic (i.e.  the correlator `looks' negative).
This is just an example of the signal to noise problem.

\subsection{Hamiltonain automatic differentiation (HAD)}
\label{sec:hamiltonain-approach}

Following the discussion in section~\ref{sec:transf-hamilt-flows} we
consider the equations of motion of the action with the added source,
Eq.~(\ref{eq:action_source}), with $\tilde J = 0 + \epsilon$
\begin{subequations}
  \label{eq:lp4eompt}
  \begin{align}
  \dot{\tilde \phi}(x) &= \tilde \pi(x)\,, \\
  \dot{\tilde \pi}(x) &= \frac{1}{2}\sum_\mu \left[ \tilde \phi(x+\mu) + \tilde \phi(x-\mu) \right] \\
  \nonumber
                        &-(4+\hat m^2)\tilde \phi(x) - 4\lambda\tilde\phi^3(x)\\
  \nonumber
  &+ \epsilon \delta_{x_0,0}\,.
  \end{align}
\end{subequations}
where both the field and momentum variables are truncated polynomials
to first order (i.e.
$\tilde \phi(x) = \phi_0(x) + \phi_1(x)\,\epsilon$).
Eqs.~(\ref{eq:lp4eompt}) can be solved order by order in $\epsilon$
using any explicit scheme.  We use the fourth order symplectic
integrator of ref.\cite{DallaBrida:2017tru}.  This integrator is very
precise and with a moderate cost, the results do not depend on the
step size.  This point is crucial, since we cannot perform an
accept-reject step.

Once we have the samples generated by the e.o.m,
$\tilde \phi^{(\alpha)} = \phi^{(\alpha)}_0 +\phi^{(\alpha)}_1\epsilon$,
the correlator is given by the first order of the (one-point)
expectation value of the field
\begin{equation}
  C(x_0) = \frac{1}{N_{\rm conf}}\sum_{\alpha = 1}^{N_{\rm conf}}\sum_{\vec x}\phi_1^{(\alpha)}(x)\,.
\end{equation}

\begin{figure}
  \centering
  \includegraphics[width=0.49\textwidth]{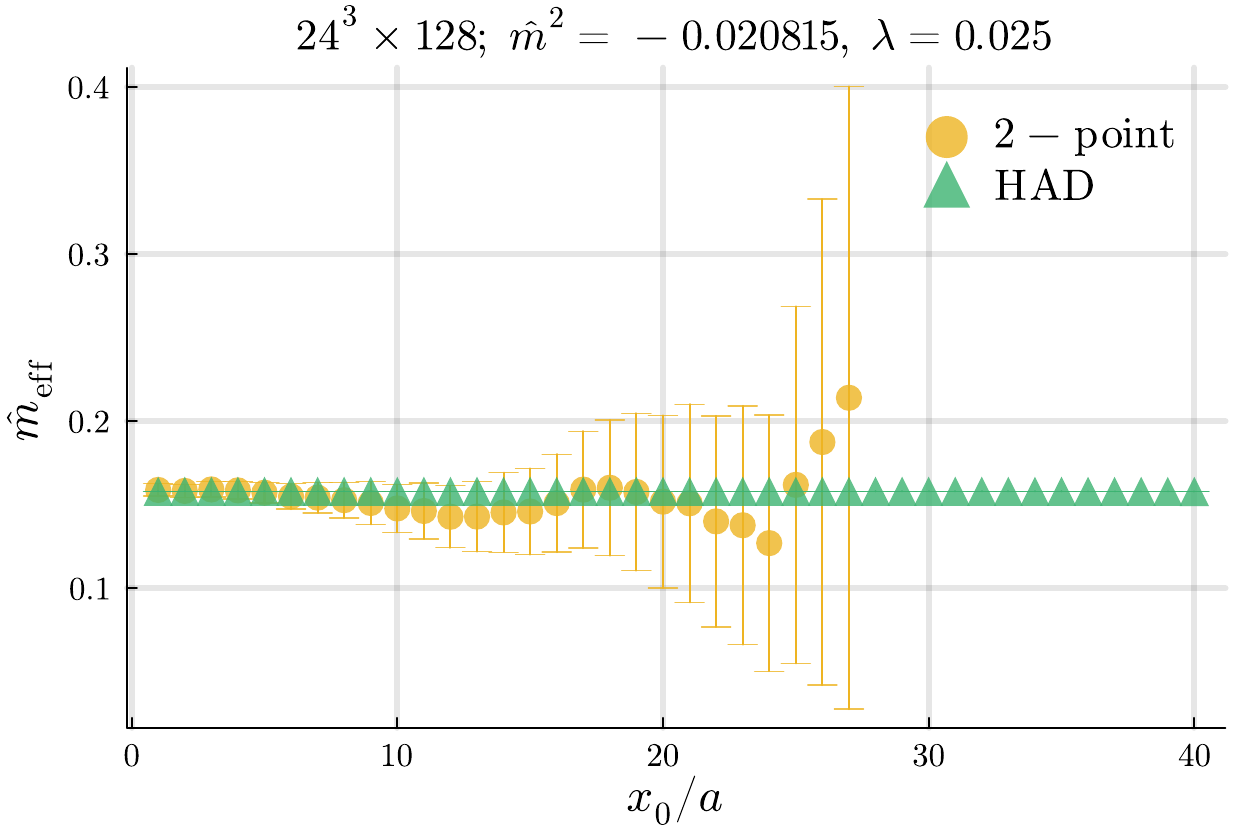}
  \caption{Effective mass Eq.~(\ref{eq:meff}) determined using the
    Hamiltonian approach, compared with the usual computation.
  The result is very precise, and the error is independent of the
  value of the Euclidean time.}
  \label{fig:meff}
\end{figure}

Figure~\ref{fig:corr} shows in green a typical result using the
Hamiltonian method (labelled HAD).  Comparing with the usual
determination based on the two-point function, in this case we observe
a constant relative error in the values of the correlator.  A more
precise statement comes from studying the effective mass
\begin{equation}
  \label{eq:meff}
  \hat m_{\rm eff} = (am_{\rm eff})(x_0) =  -\log \frac{C(x_0+a)}{C(x_0)}\,.
\end{equation}
Figure~\ref{fig:meff} shows that its error is constant, and
independent on the Euclidean time. 
Also the result is very precise: the effective mass is determined with
a precision better than 0.01\%, with a very modest statistics and
without any need to model the correlator at small Euclidean times.

\subsection{Approximate field transformations}
\label{sec:appr-field-transf}

\begin{figure}[t!]
  \centering
  \includegraphics[width=0.49\textwidth]{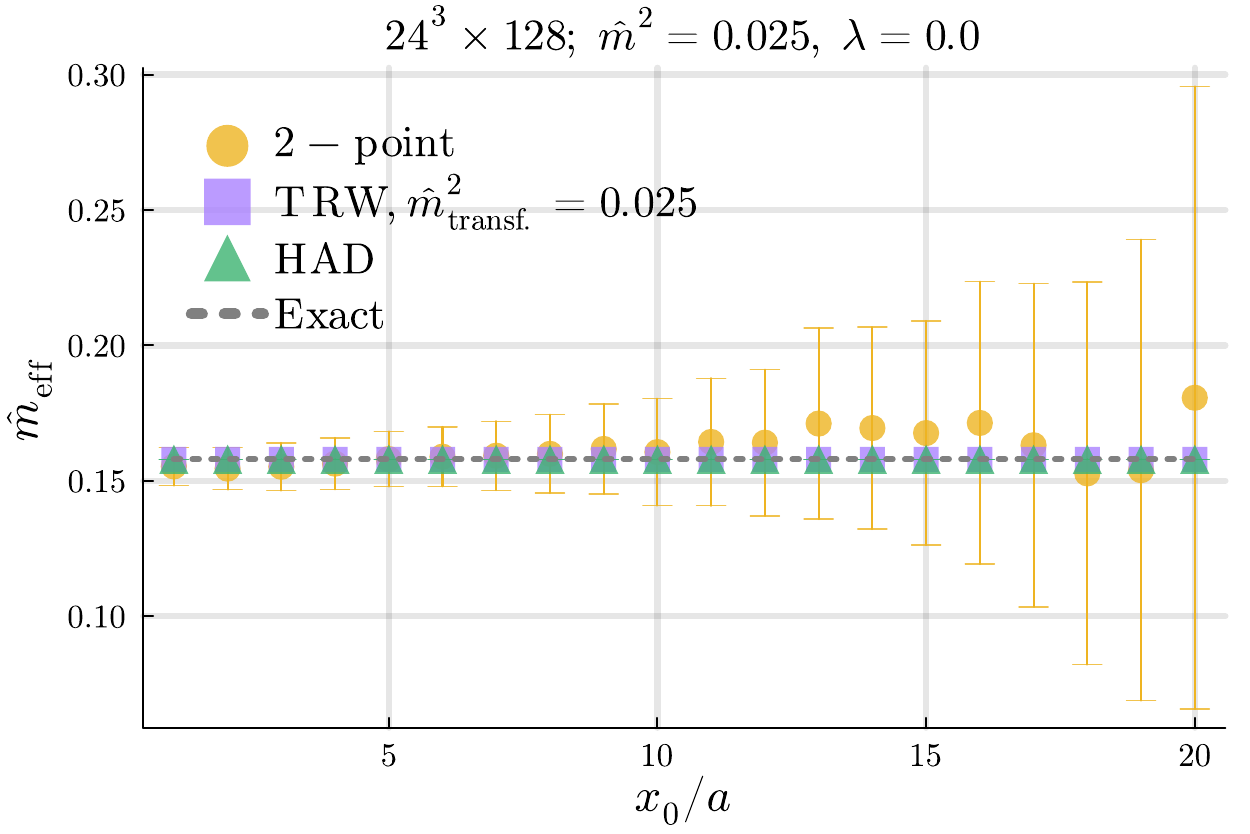}
  \caption{Free theory effective mass computed using the usual 2-point
correlator, the transformed reweighing and the Hamiltonian approach. Since the
transformation is exact for $\lambda=0$ the TRW and HAD methods give similar and
very precise results. The exact lattice free theory mass is shown for
comparison.}
  \label{fig:meff_freetheory}
\end{figure}

Although finding analitically a field transformation that obeys
exactly Eq.~(\ref{eq:exact}) is in general very difficult, a
perturbative approach might be sucessfull. In fact, if one uses a
Fourier representation of the field
\begin{equation}
  \phi_p = \sum_{x} e^{\imath p x} \phi(x)\,,
\end{equation}
the action reads
\begin{equation}
  S[\phi] = \sum_p \phi_p^*(\hat p^2 + \hat m^2)\phi_p + \mathcal O(\lambda)\,,
\end{equation}
with
\begin{equation}
  \hat p_\mu = \frac{2}{a}\sin ap_\mu\,,\quad \left(\hat p^2 = \sum_\mu\hat p_\mu\hat p_\mu\right)\,.
\end{equation}
It is straightforward to check that the transformation
\begin{equation}
  \label{eq:transl0}
  \phi_p \to \tilde \phi_p = \phi_p + \frac{1}{2}\frac{\delta(\vec p)}{\hat p^2 + \hat m^2}\, \epsilon
\end{equation}
is an exact transformation for the free case (i.e. it obeys
Eq.~(\ref{eq:exact}) when $\lambda=0$).
Figure~\ref{fig:meff_freetheory} shows that this transformation
(labeled TRW) solves the signal to noise problem: the error in the
effective mass is now independent of $x_0$. This supports our point of
view: \emph{the root of the signal to noise problem is in the large
fluctuations of the reweighting factors}. Once they are made constant by a
suitable field transformation, the signal to noise problem disappears.

At $\lambda\ne 0$ the transformation Eq.~(\ref{eq:transl0}) is no
longer exact, but one can still use the transformation if one includes
the appropriate reweighting factors Eq.~(\ref{eq:trw_factors}).
Due to the additive mass renormalization
characteristic of scalar field theories, keeping the renormalized mass
constant in lattice units as one increases the interaction requires to
use negative squared bare masses ($\hat m^2 < 0$). In these cases the
naive transformation Eq.~(\ref{eq:transl0}) does not capture the main
physical properties of the system, and the transformation
Eq.~(\ref{eq:transl0}) does not reduce the signal to noise problem
except for extremely small values of the interaction.

\begin{figure}[htb!]
  \centering
  \includegraphics[width=0.49\textwidth]{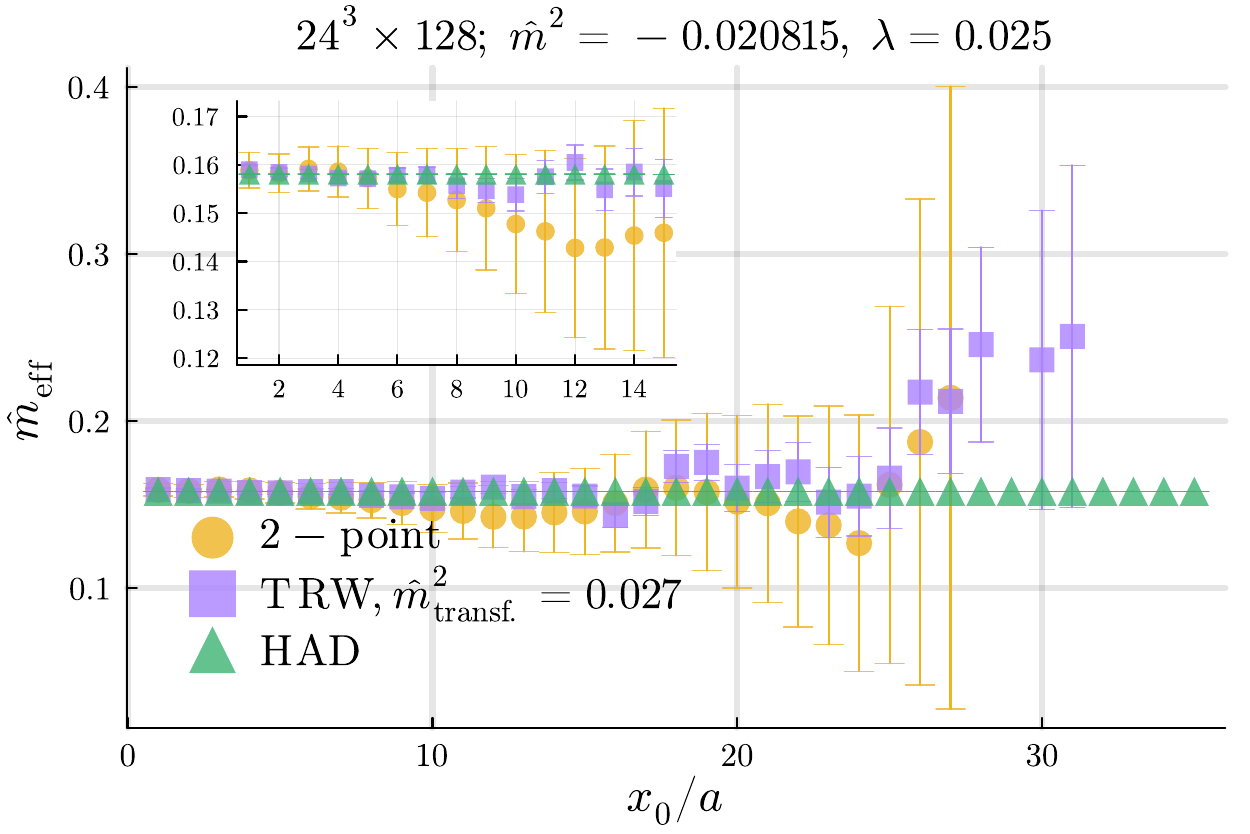}
  \caption{Effective mass computed using the usual 2-point correlator, the
transformed reweighing and the Hamiltonian approach. The bare couplings,
$\hat m^{2}=-0.020815,~\lambda=0.025$ were tuned to have a renormalized mass
$am_{R}^{2}\approx 0.025$.}
  \label{fig:meff_lam0.025}
\end{figure}

\begin{figure*}[htb!]
  \centering
  \includegraphics[width=0.49\textwidth]{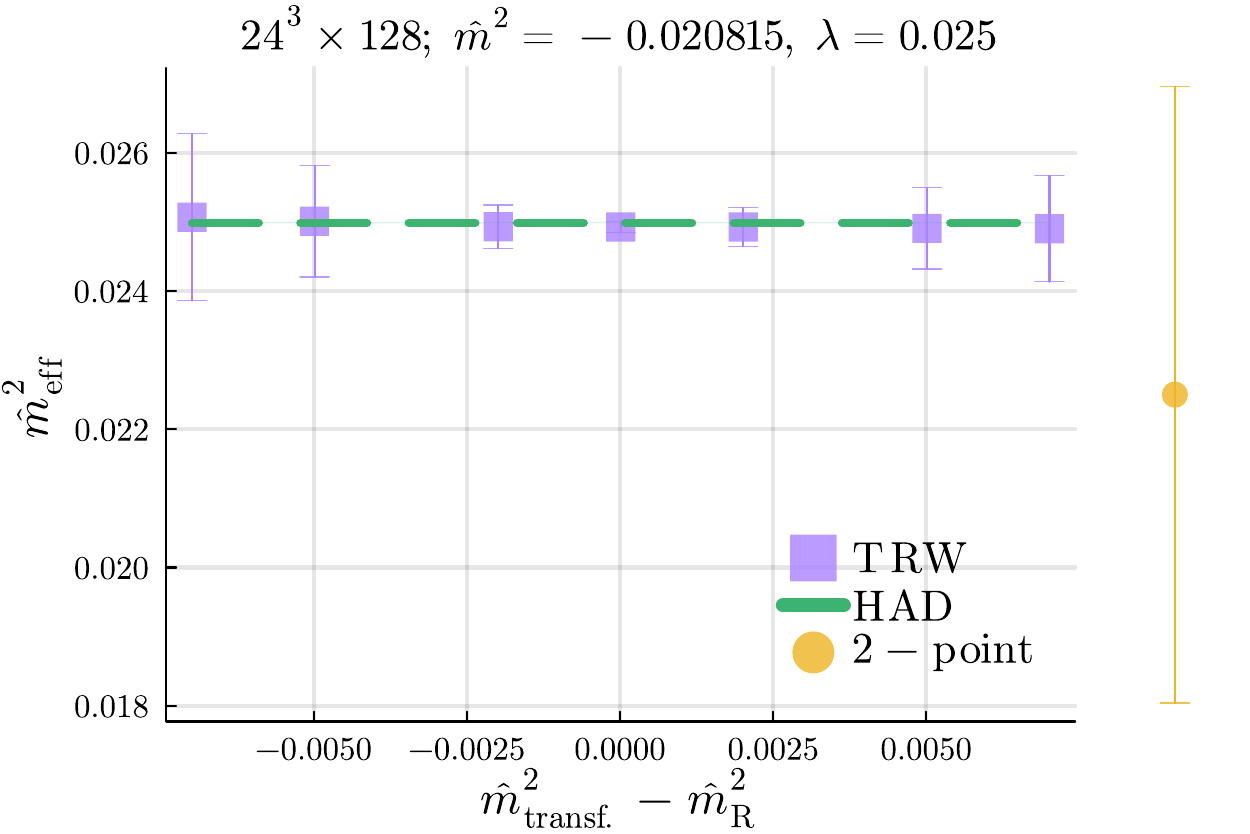}
  \includegraphics[width=0.49\textwidth]{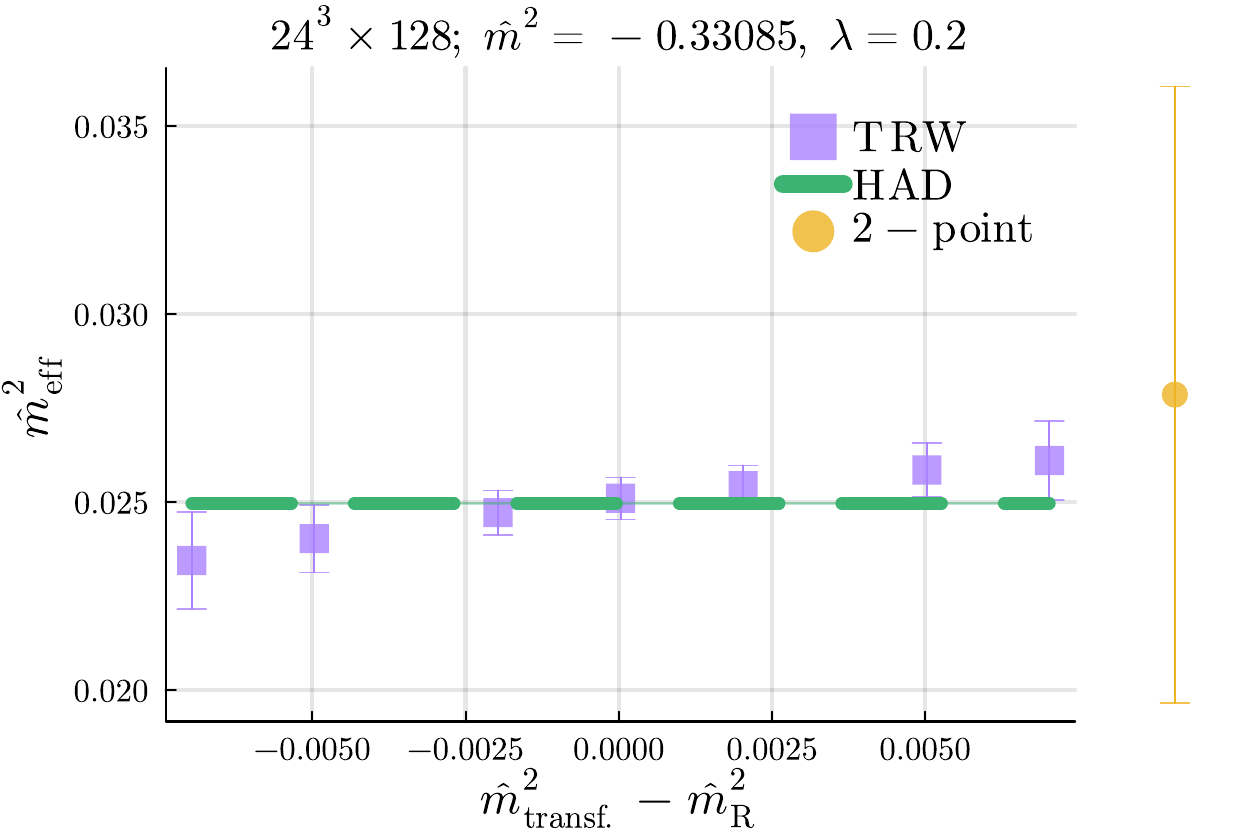}
  \includegraphics[width=0.49\textwidth]{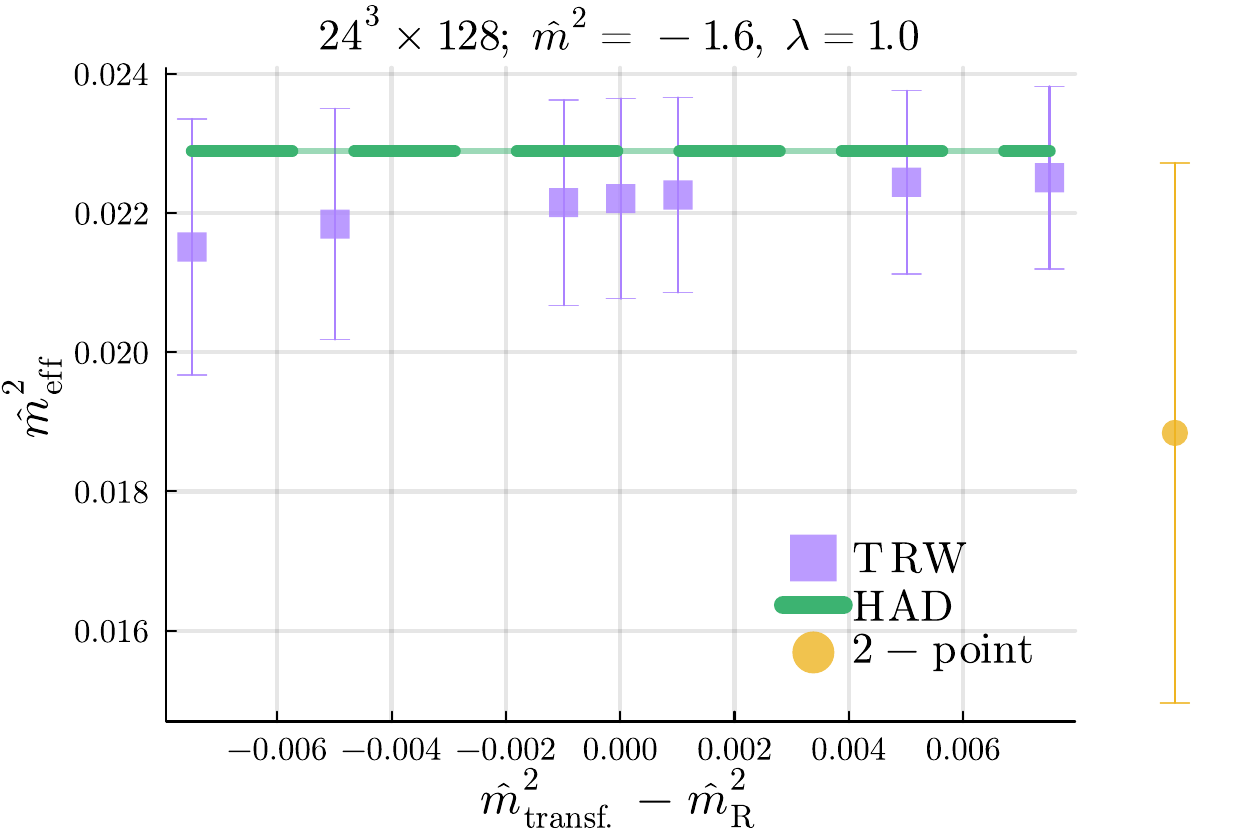}
  \caption{Fits to a constant of the effective mass squared
$m^{2}_{\textrm{eff}}$ for time slices $3<x_0 <30$.  We use different
values of $m^2_\textrm{transf.}$, in the function $f(p)$,
Eq.~(\ref{eq:transl0}), for the three values of the bare quartic
coupling $\lambda=0.025$, $\lambda=0.2$, and $\lambda=1.0$.  The
renormalized masses obtained through HAD are
$m_{\rm HAD}=0.158056(10),~0.158135(85)$, and $0.15140(15)$,
respectively. }
  \label{fig:scan}
\end{figure*}

Motivated by the presence of this additive renormalization in scalar field
theories, one can propose a simple modification to the naive
transformation, where the mass does not need to correspond with the
bare mass parameter of the Lagrangian
\begin{equation}
  \label{eq:transl0}
  \phi_p \to \tilde \phi_p = \phi_p + \frac{1}{2}\frac{\delta(\vec p)}{\hat p^2 + \hat m_{\rm transf}^2}\, \epsilon\,.
\end{equation}
If $\hat m_{\rm transf}^2$ and  $\hat m_R^2$ are similar the signal to
noise problem is severely reduced, as shown in
figure~\ref{fig:meff_lam0.025}. Here we used $\lambda=0.025$ and
$\hat m_{\rm transf}^2 = 0.027$ (in this case
$\hat m_{\rm R}^2=0.025$).  A more general picture can be seen in
figure~\ref{fig:scan}, where different values of $\hat m_{\rm transf}$
are probed and we show a fit of the effective mass to a
constant for the time slices $3< x_0/a < 30$.
Even a rough approximation to the renormalized mass is
enough to produce  a significant improvement over the naive
computation of the two point function.  This conclusion holds even for
large values of the coupling (i.e.  $\lambda=1.0$, shown in the
bottom plot of Figure~\ref{fig:scan}).  Table~\ref{tab:results} allows
a more quantitative statement.  Improvements between a factor 2 and 20
are achievable with a rough approximation of the renormalized mass
(10\% approximation), with the exact improvement depending on the
value of the self coupling $\lambda$.


%

\begin{table*}
  \centering
  \begin{tabular}{lllllll}
  \toprule
    &&\multicolumn{5}{c}{$x_0$} \\
    \cmidrule(lr){3-7}
    $\lambda$ & Method & 0 & 5 & 10 & 15 & 20 \\
    \midrule
    \multirow{3}{*}{0.025} & 2-point&        0.157(16)&        0.132(19)&        0.217(61) &         0.19(12) &         0.13(67) \\
    & TRW ($\hat m_{\rm trans}^2 = 0.025$) &      0.15865(64) &       0.1577(10) &       0.1572(23) &       0.1511(49) &        0.151(11) \\
    & HAD &  0.158094(19) &     0.158087(15) &     0.158056(16) &     0.158047(15) &     0.158058(18)\\
    \midrule
    \multirow{3}{*}{0.2} & 2-point &        0.186(32) &        0.167(51) &        0.184(77) &         0.02(17) &         0.46(74) \\
    & TRW ($\hat m_{\rm trans}^2 = 0.025$) &       0.1610(49) &       0.1563(75) &        0.167(16) &        0.124(35) &        0.111(60) \\
    & HAD &      0.15798(19) &      0.15792(15) &      0.15805(16) &      0.15820(20) &      0.15804(23) \\
    \midrule
    \multirow{3}{*}{1.0} & 2-point &        0.158(25) &        0.142(27) &        0.130(49) &         0.24(14) & n.a. \\
    & TRW ($\hat m_{\rm trans}^2 = 0.0229$) &        0.165(12) &        0.142(18) &        0.130(35) &        0.137(69) &         0.32(22) \\
    & HAD &      0.15145(53) &      0.15141(32) &      0.15145(35) &      0.15130(35) &      0.15102(33) \\
  \bottomrule
  \end{tabular}
  \caption{Effective masses in lattice units (Eq.~(\ref{eq:meff}))
    obtained at different values of $x_0$ and using different methods.
The label ``2-point'' corresponds to the usual computation of the
correlator Eq.~(\ref{eq:corr}), that suffers from the signal to noise
problem: errors in the effective masses increase with $x_0$.  The
label TRW corresponds to reweighing after having done the free-field
inspired transformation Eq.~(\ref{eq:transl0}).  Here we show the case
when $\hat m_{\rm trans} \approx \hat m_{\rm R}$, that shows a
significant reduction of the error (see Figures~\ref{fig:scan}). This
reduction is more evident at small coupling (as expected, see
section~\ref{sec:appr-field-transf}), but it is significant at large
values of the coupling.  Finally HAD labels the Hamiltonian method
(section~\ref{sec:hamiltonain-approach}), where the signal to noise
problem is completely absent, and the error in the effective masses
are basically independent on $x_0$.}
  \label{tab:results}
\end{table*}

At this point it is important to refer that contrarily to the
Hamiltonian method, reweighting does not require a dedicated
simulation. 
Scanning over different values of $m_{\rm trans}$ is numerically
cheap. Moreover, the issues with 
the convergence of the stochastic process (cf.
section~\ref{sec:transf-hamilt-flows}) are not present in this
approach.

\section{Conclusions}
\label{sec:conclusions}

The signal to noise problem is ubiquitous in lattice field theory calculations.
Correlators quickly lose their signal at large Euclidean times. This is
particularly problematic because this is the crucial regime to extract many
physical properties (i.e.  the spectrum, form factors, or the anomalous magnetic
moment of the muon, to name a few examples).

This work shows that a \emph{solution} to the signal to Noise problem
exists. 
It comes from determining  correlators not as $n-$point functions,
but \emph{as derivatives of one-point functions with repect to sources added
to the action}. 
The framework of Stochastic Automatic Differentiation
(SAD)~\cite{Catumba:2023ulz} allows to understand that these
derivatives can be numerically estimated in many different ways, with
very different variances. 
Generically the numerical computation of a correlator has to be
understood as a reweighing calculation, with the usual $n-$point
function estimate being equivalent to a naive application of the
reweighing formula. 
The large variance that produces the signal to noise problem comes
from the statistical fluctuations of the reweighing factors. 
Change of variables that reduce the variance of reweighing factors
allows to alleviate the signal to noise problem. This point of view is
complementary to the usual arguments based on the spectral
decomposition of the variance of correlators~\cite{Parisi:1983ae,Lepage:1989hd}.

As we have seen using a scalar theory as toy model, when the change of
variables lead to constant reweighing factors, the signal to noise
problem is completely solved: correlators show a constant relative
error, independent of the Euclidean time.  Hamiltonian dynamics is
able to generate these exact change of variables for some models.
Unfortunately, the Hamiltonian approach is not generically applicable,
and in particular to employ this method for QCD one would need to
extend the Hamiltonian framework to theories with compact variables,
and to deal with non-real terms in the action.  On the other hand we
have shown that approximate field transformations can still lead to
significant improvements: in the scalar field theory example a
free-field inspired transformation produces significant reductions in
the error of the correlator,  even at large values of the self
coupling $\lambda$.

For the case of gauge theories or QCD, the lattice community has
gained a lot of experience studying several type of field
transformations: \emph{trivializing
  maps}~\cite{Luscher:2009eq,Bacchio:2022vje} and
\emph{normalizing flows} (see for
example~\cite{Albergo:2021vyo,Caselle:2022esc} and references
therein) have been used to accelerate the generation of gauge
ensembles, and similar approaches have been used to compute hadronic
matrix elements~\cite{Batelaan:2023jqp,Abbott:2024kfc} or determine
derivatives of the action with respect to the bare parameters~\cite{Bacchio:2023all}. 
It seems worth studying this transformations as a tool to tame the
signal to noise problem.
Contrary to the usual application of normalizing flows, where the
method has to beat state of the art algorithms (like the HMC), in this
case the common computation of the correlator using the two point
function definition Eq.~(\ref{eq:corr}) can only be improved by using
field transformations.

\section*{Aknowledgements}

AR thanks M. 
Peardon for the extremely useful discussions and P. 
Hernández for her comments on the manuscript. 
We acknowledge support from the Generalitat
Valenciana grant PROMETEO/2019/083, the European projects
H2020-MSCA-ITN-2019//860881-HIDDeN and 101086085-ASYMMETRY, and the national
project PID2020-113644GB-I00 as well as the technical support provided by the
Instituto de Física Corpuscular, IFIC (CSIC-UV). The authors
acknowledge financial support from the Generalitat Valenciana grant
CIDEGENT/2019/040.  The computations were performed on the local SOM clusters,
funded by the MCIU with funding from the European Union NextGenerationEU
(PRTR-C17.I01) and Generalitat Valenciana, ASFAE/2022/020. We also acknowledge
the computational resources provided by Finis Terrae II (CESGA). The
authors also gratefully acknowledge the computer 
resources at Artemisa, funded by the European Union ERDF and Comunitat
Valenciana, as well as the technical support provided by the Instituto de Física
Corpuscular, IFIC (CSIC-UV).

\bibliography{/home/alberto/docs/bib/campos.bib}

\end{document}